\documentclass[preprint2]{aastex6}

\usepackage{xspace}  
\usepackage[T1]{fontenc}
\usepackage{ifthen}
\usepackage{xstring} 
\usepackage{xifthen} 
\usepackage{array}

\newcommand{\Kepler}{\textit{Kepler}\xspace} 
\newcommand{\Spitzer}{\textit{Spitzer}\xspace} 
\newcommand{\ktwo}{\textit{K2}\xspace}
\newcommand{\terra}{\texttt{TERRA}\xspace}
\newcommand{\ReaMatch}{\texttt{ReaMatch}\xspace}
\newcommand{\Kp}{\textit{Kp}\xspace}

\newcommand{\isochrones}{\texttt{isochrones}\xspace}
\newcommand{\isoclassify}{\texttt{isoclassify}\xspace}

\newcommand{\Mstar}{\ensuremath{M_{\star}}\xspace}
\newcommand{\Rstar}{\ensuremath{R_{\star}}\xspace} 
 
\newcommand{\fe}{\ensuremath{\mathrm{[Fe/H]}}\xspace}
\newcommand{\teff}{\ensuremath{T_{\mathrm{eff}}}\xspace}  
\newcommand{\logg}{\ensuremath{\log g}\xspace} 
\newcommand{\vsini}{\ensuremath{v \sin i}\xspace} 
 
\newcommand{\jmk}{\ensuremath{J\mathrm{-}K}\xspace} 
\newcommand{\kepmag}{\ensuremath{Kp}\xspace}

\newcommand{\Rp}{\ensuremath{R_P}\xspace}

\newcommand{\kms}{km s$^{-1}$\xspace}

\renewcommand{\Re}{\ensuremath{R_{\oplus}}\xspace}

\newcommand{\Rsun}{\ensuremath{R_{\odot}}\xspace }
\newcommand{\Msun}{\ensuremath{M_{\odot}}\xspace}

\def\deg{\ensuremath{^{\circ}}}
\newcommand{\rrat}{\ensuremath{\Rp/\Rstar}\xspace} 

\newcommand{\argperi}[1]{\ensuremath{\ifthenelse{\isempty{#1}}{\omega_P}{\omega_{P,#1}}}\xspace}
\newcommand{\inc}[1]{\ensuremath{\ifthenelse{\isempty{#1}}{i}{i_{#1}}}\xspace}
\newcommand{\ecc}[1]{\ensuremath{\ifthenelse{\isempty{#1}}{e}{e_{#1}}}\xspace}
\newcommand{\per}[1]{\ensuremath{\ifthenelse{\isempty{#1}}{P}{P_{#1}}}\xspace}
\newcommand{\node}[1]{\ensuremath{\ifthenelse{\isempty{#1}}{\Omega}{\Omega_{#1}}}\xspace}
\newcommand{\meananom}[1]{\ensuremath{\ifthenelse{\isempty{#1}}{M}{M_{#1}}}\xspace}
\newcommand{\ecosw}[1]{\ensuremath{\ifthenelse{\isempty{#1}}{e \cos \omega_P}{e \cos \omega_{#1}}}\xspace}
\newcommand{\esinw}[1]{\ensuremath{\ifthenelse{\isempty{#1}}{e \sin \omega_P}{e \sin \omega_{#1}}}\xspace}
\newcommand{\T}[1]{\ensuremath{\ifthenelse{\isempty{#1}}{T}{T_{#1}}}\xspace}
\newcommand{\tc}[1]{\ensuremath{\ifthenelse{\isempty{#1}}{T_{c}}{T_{c,#1}}}\xspace}

\newcommand{\sigjit}[1]{
	\ensuremath{
		\ifthenelse{\equal{#1}{}}{\sigma_\mathrm{jit}}{\sigma_\mathrm{jit,#1}}}
	\xspace
}

\newcommand{\npix}{\ensuremath{N_{\mathrm{pix}}}\xspace}
\newcommand{\stat}[1]{%
\IfEqCase{#1}{%
{n-lightcurves}{87913}
{n-hires-spec}{143}
{n-hires-spec-eb}{8}
{n-hires-spec-pc}{105}
{n-hires-spec-other}{30}
{n-sm-smsyn}{116}
{n-sm-smemp}{14}
{n-sm-both}{130}
{n-sm-none}{13}
{n-k2oi}{167}
{n-k2oi-stars}{157}
{n-k2oi-eb}{16}
{n-k2oi-not-eb}{151}
{n-k2oi-not-eb-stars}{141}
{n-k2oi-not-eb-stars-1}{132}
{n-k2oi-not-eb-stars-2}{8}
{n-k2oi-not-eb-stars-3}{1}
{kepmag-k2oi-not-eb}{12.8}
{kepmag-koi}{14.6}
}[\PackageError{tree}{Undefined option to tree: #1}{}]%
}%
\newcommand{\SpecMatchSyn}{\texttt{SpecMatch-Syn}\xspace}
\newcommand{\SpecMatchEmp}{\texttt{SpecMatch-Emp}\xspace}

\begin{document}


\title{Planet Candidates from K2 Campaigns 5--8 and Follow-up Optical Spectroscopy}



\author{Erik A. Petigura\altaffilmark{1,9,10}}
\author{Ian J. M. Crossfield\altaffilmark{2,3}}
\author{Howard Isaacson\altaffilmark{4}}
\author{Charles A. Beichman\altaffilmark{5}}
\author{Jessie L. Christiansen\altaffilmark{5}}
\author{Courtney D. Dressing\altaffilmark{1,4}}
\author{Benjamin J. Fulton\altaffilmark{1,6,11,12}}	
\author{Andrew W. Howard\altaffilmark{1}}
\author{Molly R. Kosiarek\altaffilmark{3,12}}
\author{S\'ebastien L\'epine\altaffilmark{7}}
\author{Joshua E. Schlieder\altaffilmark{8}}
\author{Evan Sinukoff\altaffilmark{6,1}}
\and
\author{Samuel W. Yee\altaffilmark{1}}


\altaffiltext{1}{California Institute of Technology, Pasadena, CA, 91125, USA}
\altaffiltext{2}{Department of Physics, Massachusetts Institute of Technology, Cambridge, MA, USA}
\altaffiltext{3}{University of California Santa Cruz, Santa Cruz, CA, 95064, USA}
\altaffiltext{4}{Department of Astronomy, University of California, Berkeley, CA 94720, USA}
\altaffiltext{5}{NASA Exoplanet Science Institute \& Infrared Processing and Analysis Center, California Institute of Technology,  Pasadena, CA, 91125, USA}
\altaffiltext{6}{Institute for Astronomy, University of Hawai'i, Honolulu, HI 96822, USA}
\altaffiltext{7}{Department of Physics \& Astronomy, Georgia State University,
Atlanta, GA, 30303, USA}
\altaffiltext{8}{NASA Goddard Space Flight Center, 8800 Greenbelt Road, Greenbelt, MD 20771, USA}
\altaffiltext{9}{petigura@caltech.edu}
\altaffiltext{10}{Hubble Fellow}
\altaffiltext{11}{Texaco Fellow}
\altaffiltext{12}{NSF Graduate Research Fellow}

\begin{abstract}
We present $\stat{n-k2oi-not-eb}$ planet candidates orbiting \stat{n-k2oi-not-eb-stars} stars from \ktwo campaigns 5--8 (C5--C8), identified through a systematic search of \ktwo photometry. In addition, we identify \stat{n-k2oi-eb} targets as likely eclipsing binaries, based on their light curve morphology. We obtained follow-up optical spectra of \stat{n-hires-spec-pc}/\stat{n-k2oi-not-eb-stars} candidate host stars and \stat{n-hires-spec-eb}/\stat{n-k2oi-eb} eclipsing binaries to improve stellar properties and to identify spectroscopic binaries. Importantly, spectroscopy enables measurements of host star radii with $\approx$10\% precision, compared to $\approx$40\% precision when only broadband photometry is available. The improved stellar radii enable improved planet radii. Our curated catalog of planet candidates  provides a starting point for future efforts to confirm and characterize \ktwo discoveries.
\end{abstract}


\section{Introduction}
NASA's {\em Kepler Space Telescope}, operating in its prime mission (2009--2013; \citealt{Borucki10a}), shed light on many fundamental properties of exoplanets. Among these are the occurrence of planets as small as Earth around Sun-like and low-mass stars (e.g. \citealt{Petigura13b,Dressing15}) and the diversity of planetary bulk compositions \citep{Marcy14,Weiss14,Rogers15} extending down to Earth-size (e.g. \citealt{Howard13,Jontof-Hutter15}).

Now operating in its two-wheel \ktwo mode \citep{Howell14}, \Kepler observes a different region of sky every three months. \ktwo is conducting a wider, more shallow survey that complements the narrow, deep survey of the prime mission. Among its many accomplishments to date, \ktwo has significantly increased the number of transiting planets around moderately bright stars \citep{Crossfield16}, which will enable more detailed studies of exoplanet bulk composition using precision radial velocity facilities. \ktwo has also revealed planets around newborn stars \citep{David16,Mann16} and planets around white-dwarfs \citep{Vanderburg15WD}. Due to community-driven target selection, a large fraction of the \ktwo targets are M-dwarfs, resulting in the detection of planets in or near the habitable zone (e.g. \citealt{Crossfield15,Montet15,Petigura15,Schlieder16}).

In this paper, we provide a catalog of planet candidates and eclipsing binaries from the second year of \ktwo operations, corresponding to campaigns 5--8 (C5--C8). Section~\ref{sec:planet-candidate-search} presents our methodology for correcting spacecraft systematics in \ktwo photometry and identifying planet candidates. In Section~\ref{sec:spectroscopy}, we describe our spectroscopic follow-up program and present refined stellar parameters enabled by these spectra. We present our catalog of planet candidates and eclipsing binaries in Section~\ref{sec:planet-candidate-table} and summarize our findings in Section~\ref{sec:conclusions}.

\section{Identifying Planet Candidates}
\label{sec:planet-candidate-search}

\subsection{Photometry}
\label{sec:photometry}
During its prime mission, \Kepler achieved photometric precisions of $\approx$40~ppm on 6.5 hour timescales \citep{Christiansen12} for targets of $\approx$12~mag in the \Kepler bandpass
(i.e. $\Kp\approx12$~mag). For many stars, photometric precision was limited by intrinsic stellar variability rather than photon-limited or instrumental errors. This exquisite precision was due in large part to stable pointing enabled by four (and later three) reaction wheels which stabilized the telescope against solar radiation pressure across the three axes of the telescope. Photometry for individual target stars was extracted using stationary software apertures composed of integer numbers of connected pixels. 

During \ktwo operations, where the spacecraft uses the two remaining operational reaction wheels, solar radiation pressure causes drifts of $\sim$1~pixel to occur on $\sim$6~hr timescales. As stars drift across the CCD, variations in pixel sensitivities and variable aperture losses result in apparent brightness variations. 

Several techniques have been developed to correct for the position-dependent brightness variations due to the unstable platform of \ktwo. A non-exhaustive list includes {\tt k2sff} \citep{Vanderburg14}, {\tt k2phot} \citep{Crossfield15,Petigura15,Crossfield16}, and {\tt k2sc} \citep{Aigrain15} which model stellar brightness as a function of spacecraft orientation with a function and remove it from the light curve. The {\tt everest} package \citep{Luger16} builds on the pixel-level decorrelation (PLD) approach developed for \Spitzer \citep{Deming15} and decorrelates against the pixel-by-pixel photometric timeseries. 

We generated light curves for \stat{n-lightcurves} stars observed by \ktwo during C5--C8 using the publicly-available {\tt k2phot} Python package.%
\footnote{https://github.com/petigura/k2phot (commit a0d507)}
The general methodology is described in previous works \citep{Crossfield15,Petigura15,Crossfield16}. However, due to the evolving nature of \ktwo systematics, we have continued to adapt and refine {\tt k2phot} and summarize the methodology below.

Since systematics in the photometry are largely due to pointing drifts, accurate knowledge of the spacecraft orientation is important for removing orientation-dependent systematics. We characterize the time-dependent orientation of the \Kepler spacecraft by analyzing the positions of $\approx$100 bright but unsaturated stars having $\kepmag\approx12$~mag on a representative output channel of the \Kepler CCD.%
\footnote{The \Kepler CCD contains 84 output channels (Kepler Instrument Handbook; Van Cleve et al. 2016), of which 76 were operational during C5-C8. An additional module (4 output channels) failed during C10.}
For each long-cadence integration, we solve for the affine transformation that maps that frame to an arbitrary reference frame. We then use the sequence of affine transformations to transform a reference pixel coordinate on a reference frame%
\footnote{For example, a pixel on row 500, column 500, and frame 2000.}
to the pixel coordinate on all other frames in the campaign.

We extract photometry using stationary apertures. Aperture size is described by a single variable \npix, the number of pixels in the aperture. Apertures are constructed to accommodate image motion during a campaign. We construct apertures using a composite image constructed from the 90th percentile intensity value of all frames in a campaign. Because the stars move during \ktwo observations, the 90th percentile image is smeared out and the apertures constructed from this image accommodate the drifts, mitigating severe aperture losses. The apertures are then constructed by selecting the pixel closest to the expected position of the target star, as predicted by the WCS coordinates provided by the \ktwo project. Additional pixels are added iteratively by selecting the brightest pixel touching the current mask. 

During the photometric extraction, we search for the aperture size $N_\mathrm{pix,min}$ that minimizes noise on three-hour times scales. This aperture size strikes a balance between the desire to minimize systematic noise which grows with {\em decreasing} aperture size, and background noise, which grows with {\em increasing} aperture size. We select an initial size $N_\mathrm{pix,0}$, which is motivated by previous analyses of stars with similar \kepmag. We then try six logarithmically-spaced \npix between $N_\mathrm{pix,0} / 4$ and $N_\mathrm{pix,0} \times 4$, which samples a curve describing noise as a function of \npix. We find $N_\mathrm{pix,min}$ using upto three iterations of Newton's method. While testing different aperture sizes, we constrain  \npix to be between nine pixels and the total number of pixels in the target pixel file.

After extraction of the photometry we have a sequence of flux as a function of $x$, $y$, and $t$. We model out changes in flux that correlate with changes in $x$, $y$, and $t$ using a Gaussian process with a squared-exponential covariance kernel, which is characterized by the following seven hyper-parameters $A_x$, $l_x$, $A_y$, $l_y$, $A_t$, $l_t$, and $\sigma$. Here, $A$ corresponds to the amplitude of the GP, $l$ corresponds to characteristic length scale, and $\sigma$ accounts for a white noise component. Choosing the appropriate hyper-parameters can be computationally intensive on a star-by-star basis. We therefore adopt a scheme from \cite{Aigrain15} which optimizes the hyper-parameters on a subset of the photometry using a differential evolution global optimizer \citep{Storn97}.

\begin{figure*}
\gridline{\fig{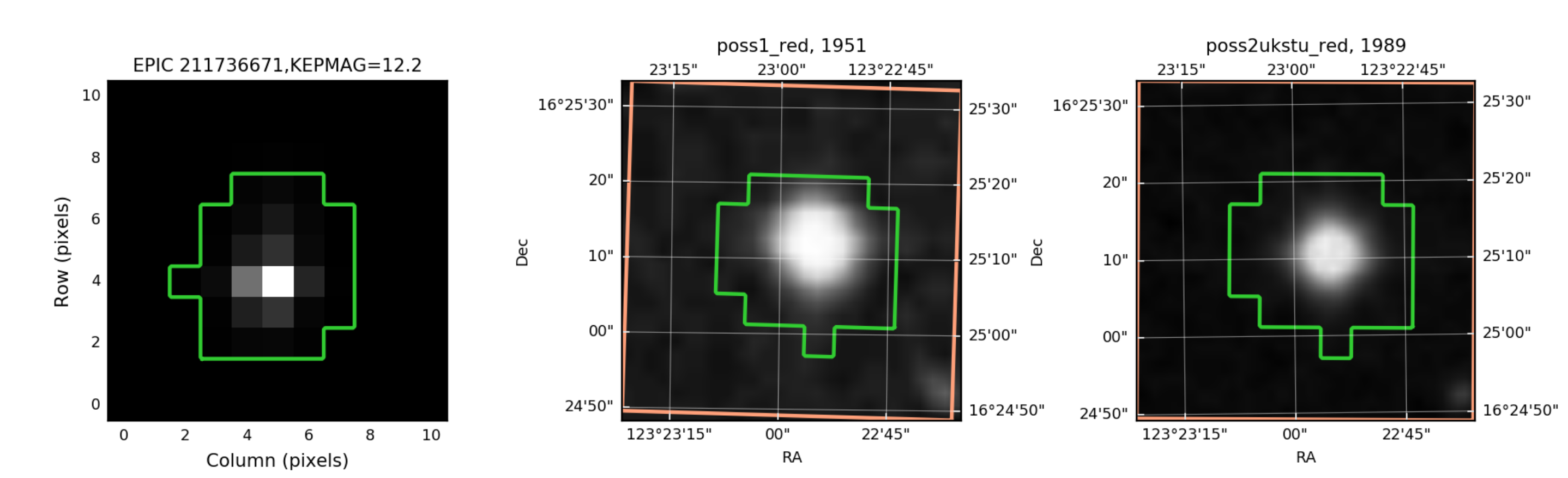}{0.9\textwidth}{(a)}}
\gridline{\fig{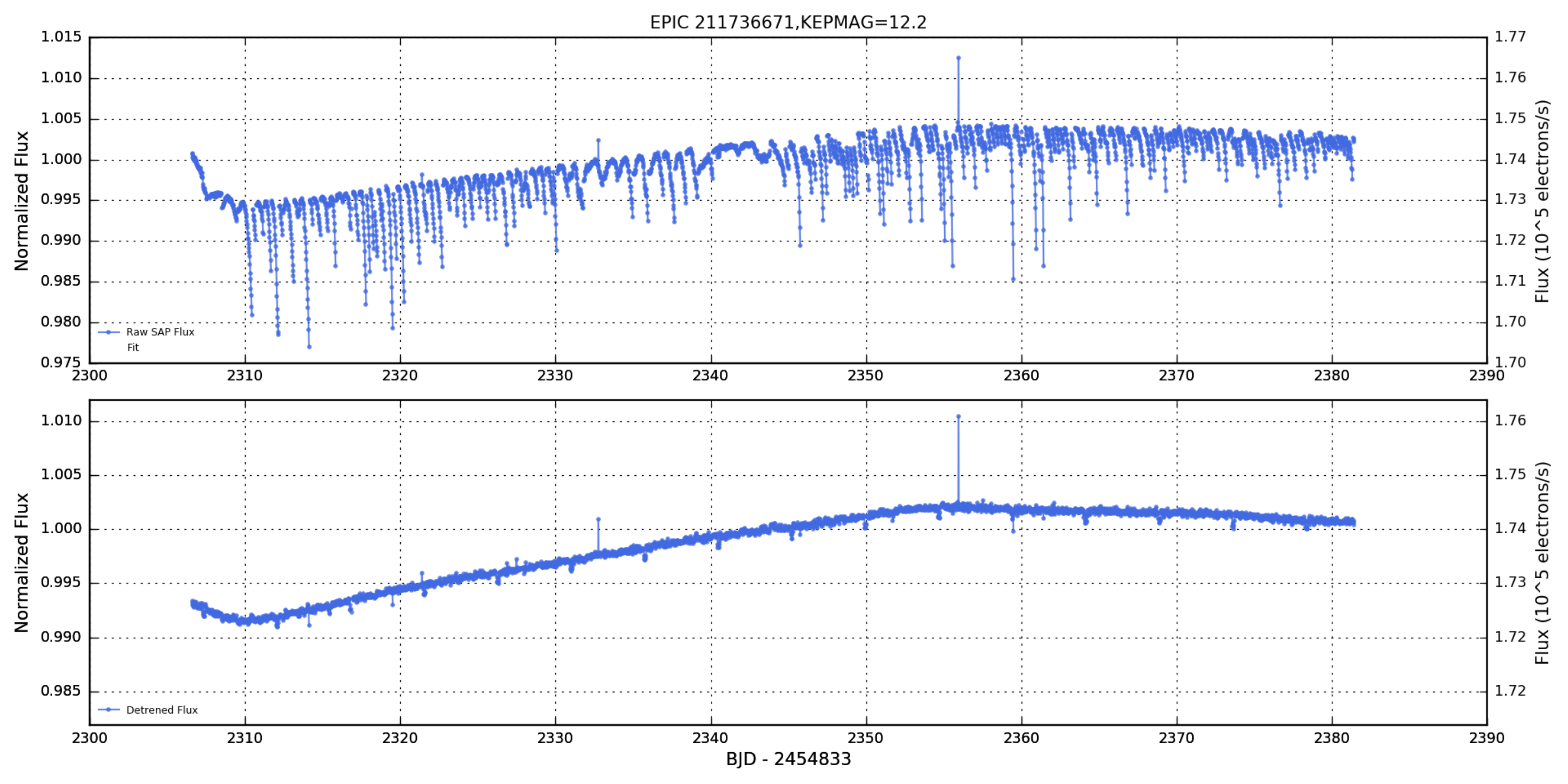}{0.9\textwidth}{(b)}}
\caption{An example of the photometric diagnostic plots included as standard data products on the ExoFOP. Panel {\bf (a)}: three images of an example planet candidate (EPIC-211736671), observed in C5. {\em Left:} the median of all long cadence C5 frames with the optimal extraction aperture shown in green. {\em Center:} the same region of sky as observed by the first Palomar Observatory Sky Survey (POSS-I). The orange region corresponds to the boundaries of the K2 frame. {\em Right:} the same region of sky as observed by POSS-II. Panel {\bf (b):} aperture photometry before and after subtraction of our systematic noise model.\label{fig:211736671}}

\end{figure*}

We produced light curves for \stat{n-lightcurves} stars in C5--C8, which are available on the Exoplanet Follow-up Observing Program (ExoFOP) website.%
\footnote{https://exofop.ipac.caltech.edu/}
Along with the photometry, we included photometric diagnostic plots showing the extraction aperture and resulting detrended light curve. Figure~\ref{fig:211736671} shows these diagnostic plots for an example planet candidate around EPIC-211736671.

\subsection{Transiting Planet Search} 
Our general transit search and vetting process is similar to that described in \cite{Crossfield16}. We give a brief summary of here. We searched the calibrated photometry for transiting planets using the publicly-available \terra algorithm.%
\footnote{https://github.com/petigura/terra (commit 9739e9)}
\terra is a matched filter-based approach and is described by \cite{Petigura13b}. \terra convolves the photometry with a box-shaped approximation of a transit profile to compute a Single Event Statistic (SES) at every \ktwo long-cadence measurement. The SES time series is phase-folded according to a finely-spaced grid of trial periods $P$ and times of first transit \T{0}.

A classical matched-filter algorithm would then compute a Multiple Event Statistic (MES) by summing SES at each trial (\per, \T{0}), which is optimal given uncorrelated Gaussian noise . However, in \ktwo photometry we observe more frequent non-Gaussian anomalies relative to \Kepler prime photometry due to the aggressive detrending that must be performed in order to remove the instrumental systematics described above. A traditional MES computation resulted in many spurious peaks with apparently high MES, but were later easily identifiable as anomalies through inspection. We address outliers by calculating MES only after removing the two highest SES peaks at each trial (\per, \T{0}).  Spurious peaks due to the chance alignment of two outliers are eliminated. This non-linear filter removes many spurious detections and eases the burden during manual vetting, described below. One consequence is that \terra does not identify planets with one or two transits occurring in a \ktwo campaign. Such transits are sometimes identified by eye, but we caution that many are likely overlooked. These events are especially amenable to searches by citizen scientists (see e.g., Christiansen et al. {\em accepted in AJ}).%
\footnote{\url{https://www.zooniverse.org/projects/ianc2/exoplanet-explorers}}

\terra identifies $\sim$1000 Threshold-Crossing Events (TCEs) per campaign. A TCE is a particular combination of (\per, \T{0}) that has MES that exceeds some threshold, but may not be an astrophysical transit. If a candidate has a periodic dimming of a consistent shape it is elevated to the status of a ``\ktwo Object of Interest'' (K2OIs) which are consistent with an astrophysical transit or eclipse. 

Our team visually inspects each K2OI to look for a robust indication that the target is an eclipsing binary. We look for secondary eclipses, which indicate that the transiting object is self-luminous and not a planet. Secondary eclipses associated with binaries with circular orbits are shifted in phase from the primary eclipse by $180^{\circ}$. We search for secondary eclipses at all phases to allow for eccentric orbits. We also look for obvious odd/even variations, which indicate that \terra has identified a nearly circular EB at half the orbital period. We also identify stars that show variability that is phase-locked to the eclipse, which is a strong indicator of star-star modulation (ellipsoidal, reflection, or relativistic beaming). 

Our eclipsing binary designation does not incorporate transit depth or whether the light curve is V-shaped. While these attributes are strong indicators of EB status, they are not conclusive. Planets transiting small M-dwarf stars can easily produce transits deeper than 1\% and short period transits may appear V-shaped due to the 30-minute sampling of \ktwo. We defer a detailed false positive calculation for a later paper.

In total, we identified \stat{n-k2oi} K2OIs, associated with $\stat{n-k2oi-stars}$ stars. Of these, \stat{n-k2oi-eb}/\stat{n-k2oi} are likely eclipsing binaries, and we refer to the remaining \stat{n-k2oi-not-eb} as planet candidates.

\subsection{Light Curve Fitting}
We fit the calibrated photometry according to the methodology of \cite{Crossfield16}. In brief, we used the publicly-available {\tt batman} light curve code \citep{Kreidberg15} to generate model light curves which we then compared against the photometry. We first derived a maximum likelihood solution and then derived parameter uncertainties using Markov Chain Monte Carlo (MCMC).%
\footnote{using the affine-invariant sampler of \cite{Goodman10} as implemented in Python by \cite{Foreman-Mackey15}}

In our modeling, the following parameters are allowed to vary: time of first transit $T_0$, orbital period $P$, inclination $i$, scaled semi-major axis $a/\Rstar$, planet-star radius ratio \Rp/\Rstar, orbital eccentricity $e$, longitude of periastron $\omega$, linear limb-darkening coefficient $u$, fractional light curve dilution $\delta$, and the out-of-transit flux level. 

During the fitting, we adopted the following priors:

\begin{itemize}
\item {\em Period}. Gaussian prior centered on maximum likelihood $P$ having dispersion of 0.01~days.%
\footnote{We imposed weak priors on $P$ and $T_0$ to keep the MCMC walkers from jumping too far from the likelihood mode and wandering away. After performing the MCMC exploration, we verified that $P$ and $T_0$ were more tightly constrained by the photometry than by the priors. The uncertainties on $T_0$ are typically 2.5\% of the prior width (median value) and no more than 50\% the prior width. The uncertainties on P are typically 2\% of the prior width (median value) and no more than 60\% the prior width.}
\item {\em Time of transit}. Uniform prior centered on maximum likelihood $T_0$ having dispersion of $0.06 \times P$.
\item {\em Radius ratio}. Uniform prior, $\Rp/\Rstar$ = [$-1$,$+1$]. Following \cite{Eastman13}, we allow for negative $\Rp/\Rstar$ in our sampling to avoid the Lucy-Sweeney-type bias that results from treating $\Rp/\Rstar$ as a positive-definite quantity \citep{Lucy71}.
\item {\em Eccentricity}. Gaussian prior centered at $10^{-4}$ having dispersion of $10^{-3}$. This effectively restricts the orbits to circular.
\item {\em Longitude of periastron}. Uniform prior, $\omega$ = [0,$2\pi$].
\item {\em Inclination}. Uniform prior, $i$ = [$50^{\circ}$,$90^{\circ}$].
\item {\em Limb-darkening}. Gaussian prior on $u$ where the mean and dispersion are computed using the publicly-available Limb-Darkening Toolkit (LDTK; \citealt{Parviainen15}). LDTK computes the distribution of $u$ given Gaussian constraints on \teff, \logg, and \fe. For consistency, we used \teff, \logg, and \fe constrained by broadband photometry from \cite{Huber16}. While photometrically-constrained \logg and \fe are low-precision, $u$ is only weakly dependent on these parameters, and the derived transit parameters are only weakly dependent on $u$.
\item {\em Dilution}. Log-uniform prior, $\log \delta$ = [$10^{-6}$,$10^{0}$]. Our fits do not incorporate dilution constraints, so $\delta$ always reverts to the prior. We include $\delta$ so we can incorporate dilution constraints at later times.%
\footnote{The non-zero prior on $\delta$ slightly alters the derived value of $\Rp/\Rstar$. However, because the median $\delta$ is $10^{-3}$, this amounts to a fractional change in derived radius ratio of $\Delta(\Rp/\Rstar)/ (\Rp/\Rstar) \approx \delta/ 2 = 5\times10^{-4}$, which may be ignored.}
\end{itemize}

In Table~\ref{tab:candidates}, we report 1$\sigma$ credible ranges on $P$, $T_0$, $\Rp/\Rstar$, transit duration $T_{14}$, and impact parameter $b$.

\section{Spectroscopy} 
\label{sec:spectroscopy}
\subsection{Spectroscopic Follow Program}
As a part of our team's standard follow-up efforts, we obtained optical spectra of $\stat{n-hires-spec}$ C5--C8 target stars using the HIgh Resolution Echelle Spectrometer (HIRES; \citealt{Vogt94}) on the Keck-I 10~m telescope. We gathered spectra for the purpose of improving host star parameters and to place limits on the presence of companion stars with small separations through searches for spectroscopic binaries (SB2s). We aimed to obtain a spectrum of every K2OI brighter than $V = 14.0$~mag. For G stars, this limit corresponds roughly to $\kepmag = 13.6$~mag.

Table~\ref{tab:spec} lists the C5--C8 targets that we observed with HIRES, along with the results from our stellar characterization and search for spectroscopic binaries, which are described in Sections~\ref{sec:stellar-characterization} and \ref{sec:sb2}, respectively. We obtained HIRES spectra of \stat{n-hires-spec-pc}/\stat{n-k2oi-not-eb-stars} of the planet candidate host stars and for \stat{n-hires-spec-eb}/\stat{n-k2oi-eb} of the likely EBs. In addition, we observed \stat{n-hires-spec-other} other C5--C8 targets that we did not identify as candidates. These were typically observed because they were identified as planet candidates by other groups.

\begin{figure}
\centering
\includegraphics[width=0.5\textwidth]{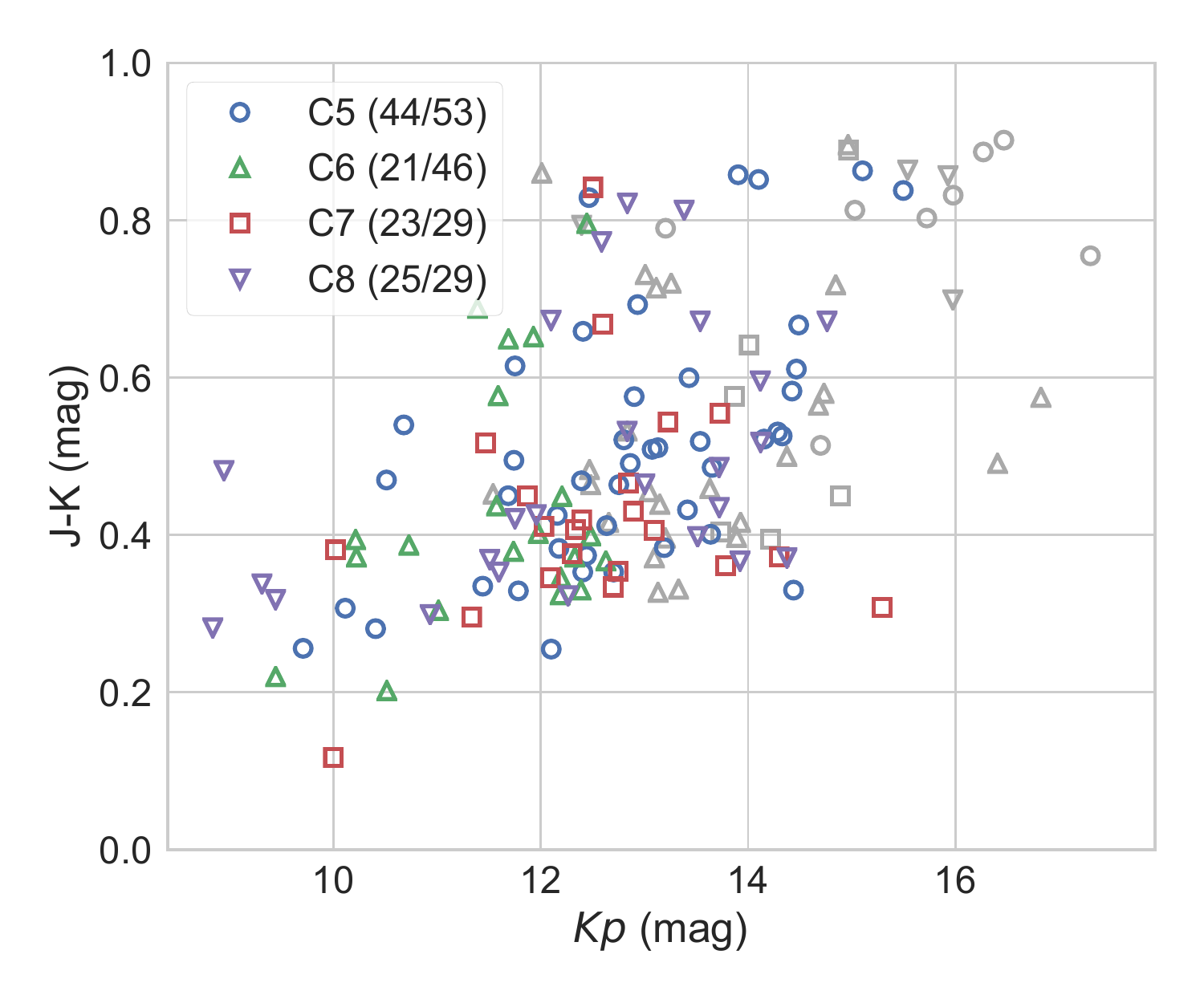}
\caption{Distribution of K2OIs as a function of \kepmag and \jmk color. The colored/gray points represent targets with/without a HIRES spectrum. The marker color and shape represent a target's specific \ktwo campaign. The HIRES follow-up is nearly complete to \Kp~=~14~mag.\label{fig:kepmag-color}}
\end{figure}

\begin{figure*}
\centering
\includegraphics[width=0.9\textwidth]{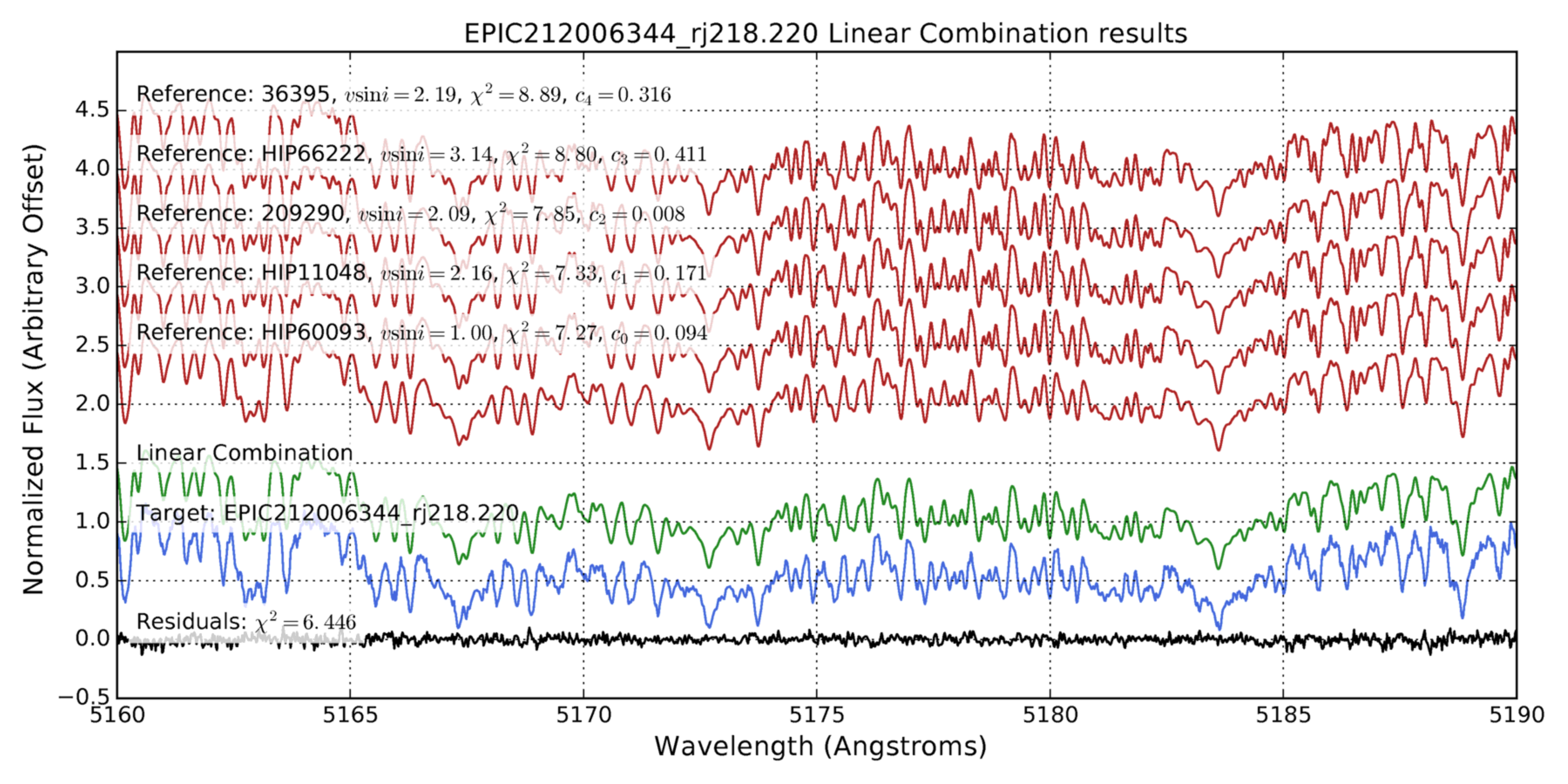}
\caption{Example HIRES characterization spectrum with \SpecMatchEmp fit. Blue spectrum is of EPIC212006344, an M0 dwarf, and illustrates the typical SNR from our characterization program of $\approx45$/pixel. The spectrum contains a dense forest of molecular lines, making {\em ab initio} spectral synthesis challenging. Red spectra are drawn from \SpecMatchEmp library and are identified as similar to the target spectrum by the \SpecMatchEmp algorithm. The green and black spectra are the best-fitting linear combination spectrum and residual spectrum, respectively.\label{fig:smemp-lincomb}}
\end{figure*}

We used the HIRES exposure meter to obtain consistent SNR levels depending on $V$-band apparent magnitude: SNR = 45 ($V < 13.0$~mag), SNR = 32 ($V = 13.0-14.0$~mag), and SNR = 20 ($V > 14.0$~mag). Exposure times ranged from $\approx$10~s for $V$ = 9~mag targets to $\approx$400~s for $V$ = 15~mag targets. SNR is computed per reduced HIRES pixel on blaze at 5500~\AA. Our HIRES follow-up was nearly complete down to $V = 14$~mag. Figure~\ref{fig:kepmag-color} shows the distribution of candidate hosts as a function of \kepmag and \jmk color. The candidates with HIRES spectra are labeled. Figure~\ref{fig:smemp-lincomb} shows a spectral segment for one K2OI to illustrate typical spectral resolution and SNR level.

\subsection{Stellar Characterization}
\label{sec:stellar-characterization}
We used our spectra to improve the precision of stellar and planetary properties such as \Rstar and \Rp. We analyzed each spectrum with one of two related publicly-available codes: \SpecMatchSyn \citep{Petigura15thesis}%
\footnote{\url{https://github.com/petigura/specmatch-syn}}
and \SpecMatchEmp \citep{Yee17}.%
\footnote{\url{https://github.com/samuelyeewl/specmatch-emp}}

\SpecMatchSyn fits five regions of optical spectrum by interpolating within a grid of model spectra from \cite{Coelho05}. Recently, \SpecMatchSyn enabled a homogeneous analysis of 1305 spectra of planet hosts identified during the prime \Kepler mission \citep{Petigura17b}. For stars with \teff = 4700--6500~K and \vsini < 20~\kms, \SpecMatchSyn achieves precisions of 60~K in \teff, 0.10~dex in \logg, and 0.04~dex in \fe, and 1~\kms in \vsini.

We converted \teff, \logg, and \fe into \Mstar and \Rstar using the publicly-available \isoclassify Python package \citep{Huber17},%
\footnote{\url{https://github.com/danxhuber/isoclassify}}
which uses the MESA Isochrones and Stellar Tracks (MIST) database \citep{Choi16,Paxton11,Paxton13,Paxton15}. While \SpecMatchSyn returns \teff precise to 60~K, as tested against other spectral synthesis codes, there are known offsets between spectroscopic \teff and other techniques such as the Infrared Flux Method (IRFM) and interferometry. For a detailed discussion of different \teff scales, see \cite{Brewer16}. To account for systematic uncertainties associated with the spectroscopic \teff scale, we have increased the \teff uncertainties to 100~K during the isochrone modeling. 

The radius uncertainties derived from \SpecMatchSyn parameters do not incorporate uncertainties associated with the MIST models themselves. \cite{Johnson17} estimated the size of these model-dependent uncertainties through a comparison of stellar radii derived using MIST models and Dartmouth Stellar Evolution Program (DSEP) models \citep{Dotter08} with identical inputs. They estimated that model-dependent radius errors are $\approx$2\% for dwarf stars ($\logg < 3.9$) and $\approx$10\% for evolved stars ($\logg>3.9$). These model-dependent uncertainties are typically smaller than the formal radius uncertainties returned by \isoclassify.

For $\teff \lesssim 4700$~K, \SpecMatchSyn does not return reliable parameters, due to the onset of molecular lines that are not well-treated in the \cite{Coelho05} models. While the high-resolution optical spectra of stars later than $\sim$K4 is challenging to compute directly, their spectra contain a wealth of information, which can be used to constrain stellar properties. \SpecMatchEmp circumvents the challenges in spectral synthesis by matching a target spectrum against an empirical spectra library of $\approx$400 touchstone stars with well-known parameters measured through other methods such as SED-fitting, interferometry, or the IRFM. \SpecMatchEmp interpolates between this library of empirical spectra to find linear combination of library spectra that best reproduces the target spectrum. \SpecMatchEmp achieves precisions of 70~K in \teff, 10\% in \Rstar, and 0.12~dex in \fe. 

\begin{figure}
\centering
\includegraphics[width=0.48\textwidth]{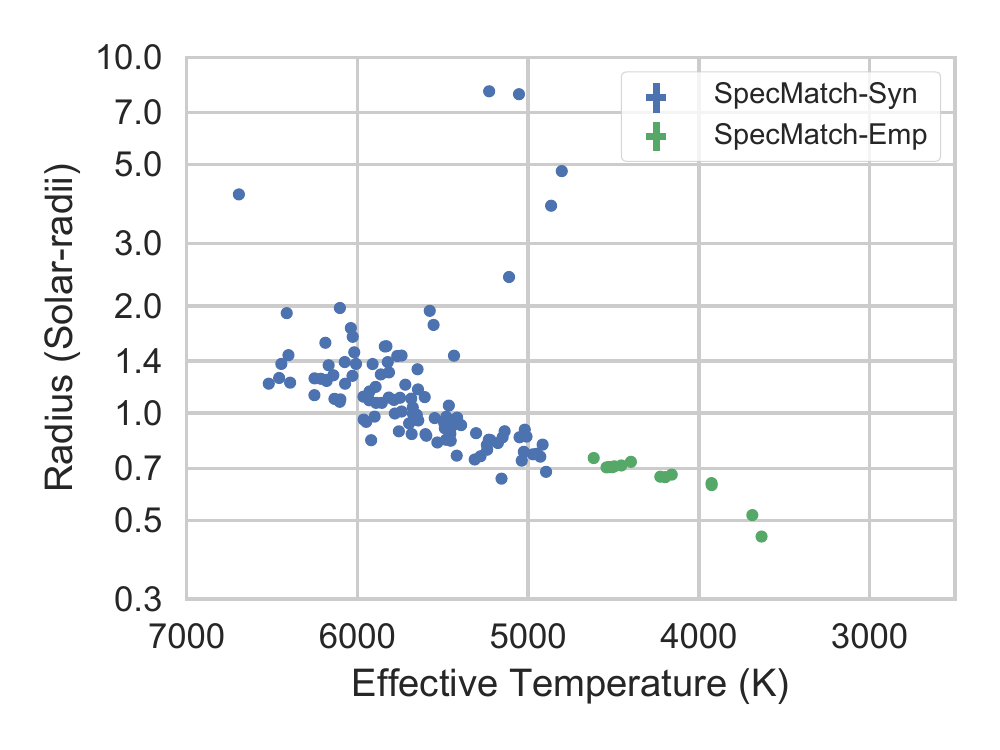}
\caption{K2OIs with reliable effective temperatures and stellar radii, as measured by \SpecMatchSyn (blue) or \SpecMatchEmp (green). Median uncertainties are shown at top right.\label{fig:spec-hr-diagram}}
\end{figure}

We adopt parameters from \SpecMatchSyn for stars hotter than 4700~K%
\footnote{as measured by \SpecMatchEmp}
and \SpecMatchEmp for cooler stars. Figure~\ref{fig:spec-hr-diagram} shows the \teff and \Rstar for K2OIs with reliable spectroscopic parameters. Our adopted stellar parameters are listed in Table~\ref{tab:spec}. Our team also conducts a parallel characterization of cool stars using NIR spectroscopy. Stellar properties up through campaign 7 are given in \cite{Martinez17} and \cite{Dressing17}.

\subsection{Searches for Spectroscopic Binaries}
\label{sec:sb2}

Each HIRES spectrum is methodically searched for secondary spectral lines using the \ReaMatch algorithm \citep{Kolbl15}.  To identify secondary spectra, each spectrum is first cross-correlated against a set of previously-observed spectra. This catalog has spectra with \teff = 3500--6500~K and \logg = 3.0--4.5~dex. The best-matching spectrum is subtracted from the target spectrum and the residuals are again cross-correlated with the catalog spectra. \ReaMatch is sensitive to companions down to 1\% the brightness of the primary having RV offsets $|\Delta v|$~>~10~\kms. Optimized for slowly-rotating FGKM stars, \ReaMatch is insensitive to SB2s orbiting primaries with \vsini > 10 \kms or with \teff outside 3500--6500 K. Table~\ref{tab:spec} lists the results of our SB2 search.

\begin{figure*}
\centering
\includegraphics[width=0.75\textwidth]{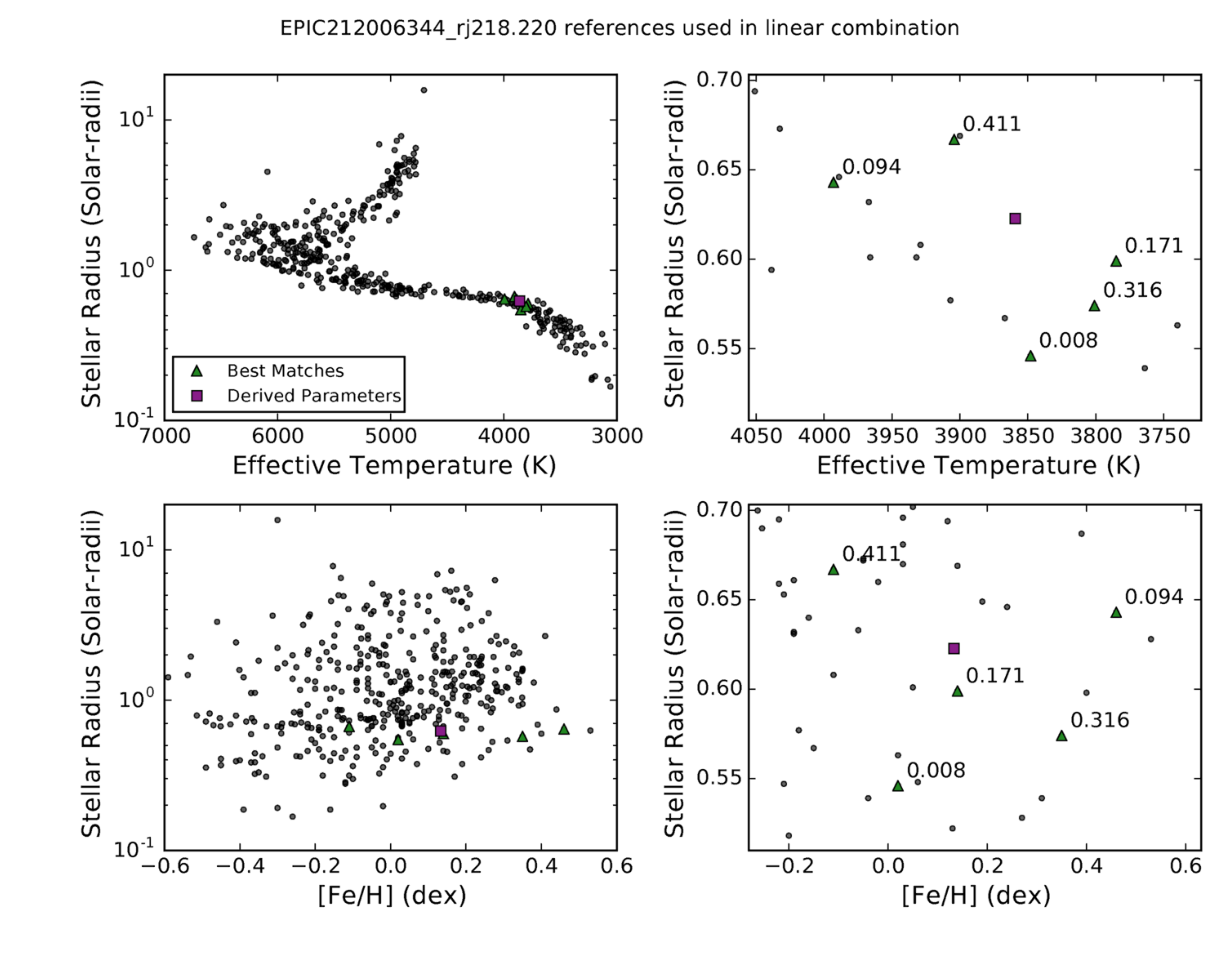}
\caption{\SpecMatchEmp characterization of EPIC212006344. Black points: \teff, \Rstar, and \fe from the \SpecMatchEmp library. Green triangles: properties of the closest-matching library spectra (red spectra in Figure~\ref{fig:smemp-lincomb}). Purple square: linear combination of library spectra spectra. The final derived parameters are \teff~=~$3925\pm70$~K, \fe = $0.43\pm0.12$~dex, \Rstar~=~$0.63\pm0.10$~dex.\label{fig:smemp-library}}
\end{figure*}

\section{Planet Candidates}
\label{sec:planet-candidate-table}

We list the \stat{n-k2oi-not-eb} planet candidates and \stat{n-k2oi-eb} likely EBs in Table~\ref{tab:candidates}. We compute planetary radius by combining \rrat measured from the light curve with the best available \Rstar. For the stars without spectra, we estimate \Rstar from broadband photometry, according to the following procedure: we estimate spectral types (SpTs) using tabulated photometric relations \citep{Kraus07,Pecaut13,Rodriguez13} and convert SpTs into \Rstar based on interferometric studies \citep{Boyajian12b}. These stellar radii are crude and we estimate their typical uncertainties to be $\approx$40\%, typical of errors derived from broadband photometry \citep{Brown11}. 

\begin{figure}
\centering
\includegraphics[width=0.5\textwidth]{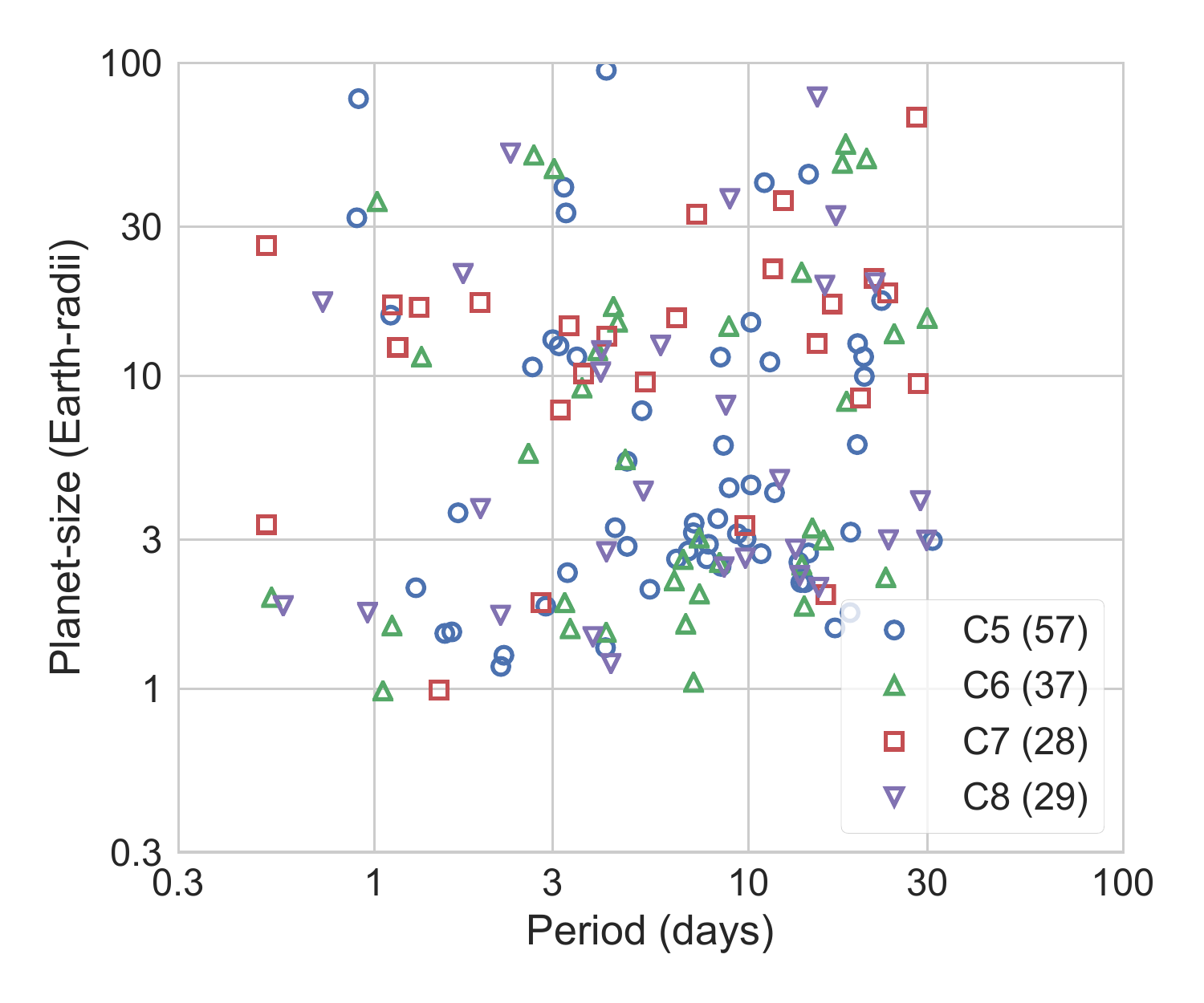}
\caption{Planet size and orbital period for $\stat{n-k2oi-not-eb}$ planet candidates identified by \ktwo in C5--C8. We have excluded targets identified as likely EBs from their light curve morphology. The legend at lower right links marker shape/color to a specific campaign and gives the total number of candidates identified in each campaign.\label{fig:per-pradius}}
\end{figure}

\begin{figure*}[ht!]
\begin{center}
\includegraphics[trim=2.cm 0.5cm 0.5cm 0cm,width=0.32\textwidth]{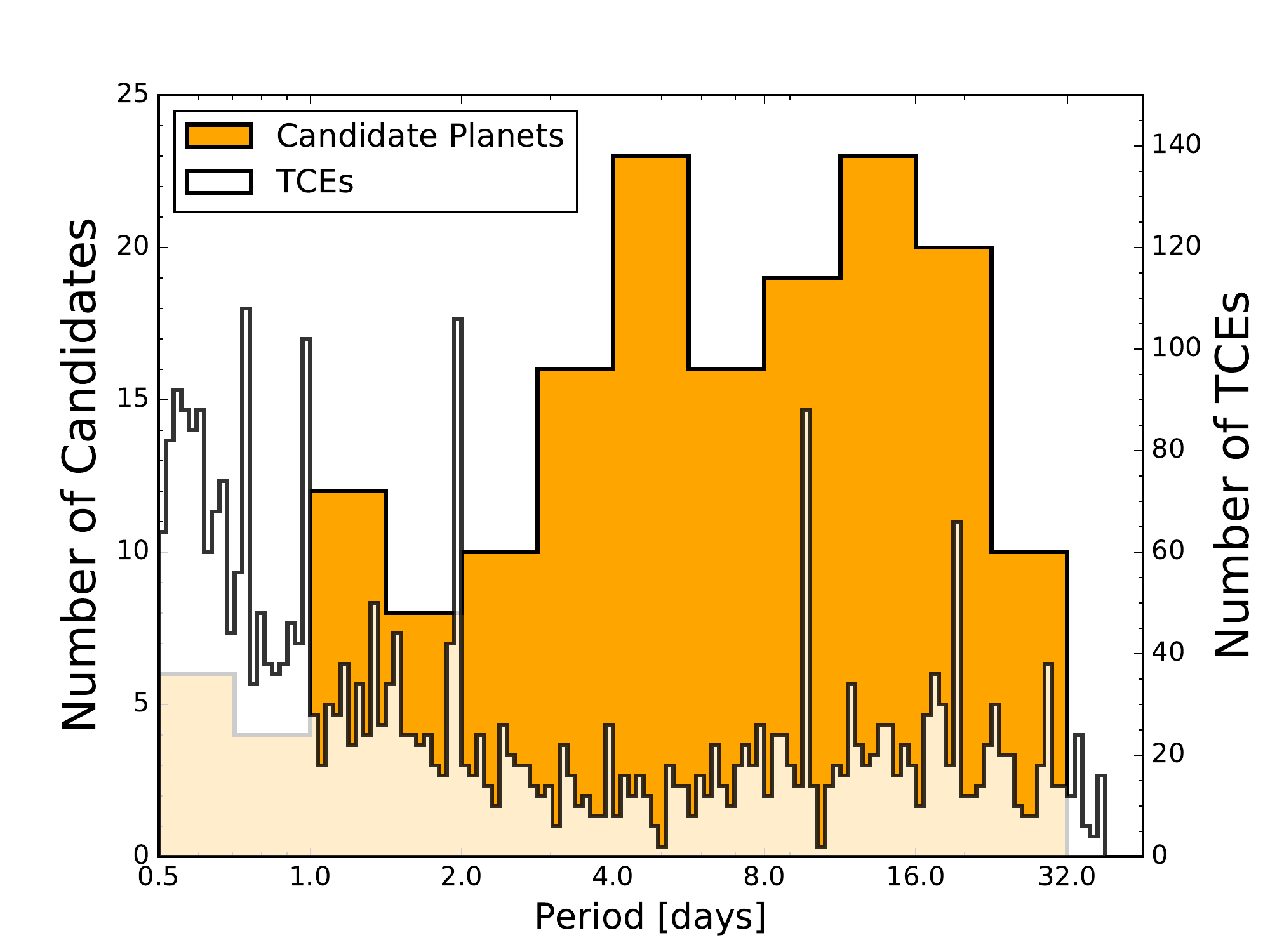}
\hspace{0.1cm}
\includegraphics[trim=0.5cm 0.5cm 1.5cm 0cm,width=0.32\textwidth]{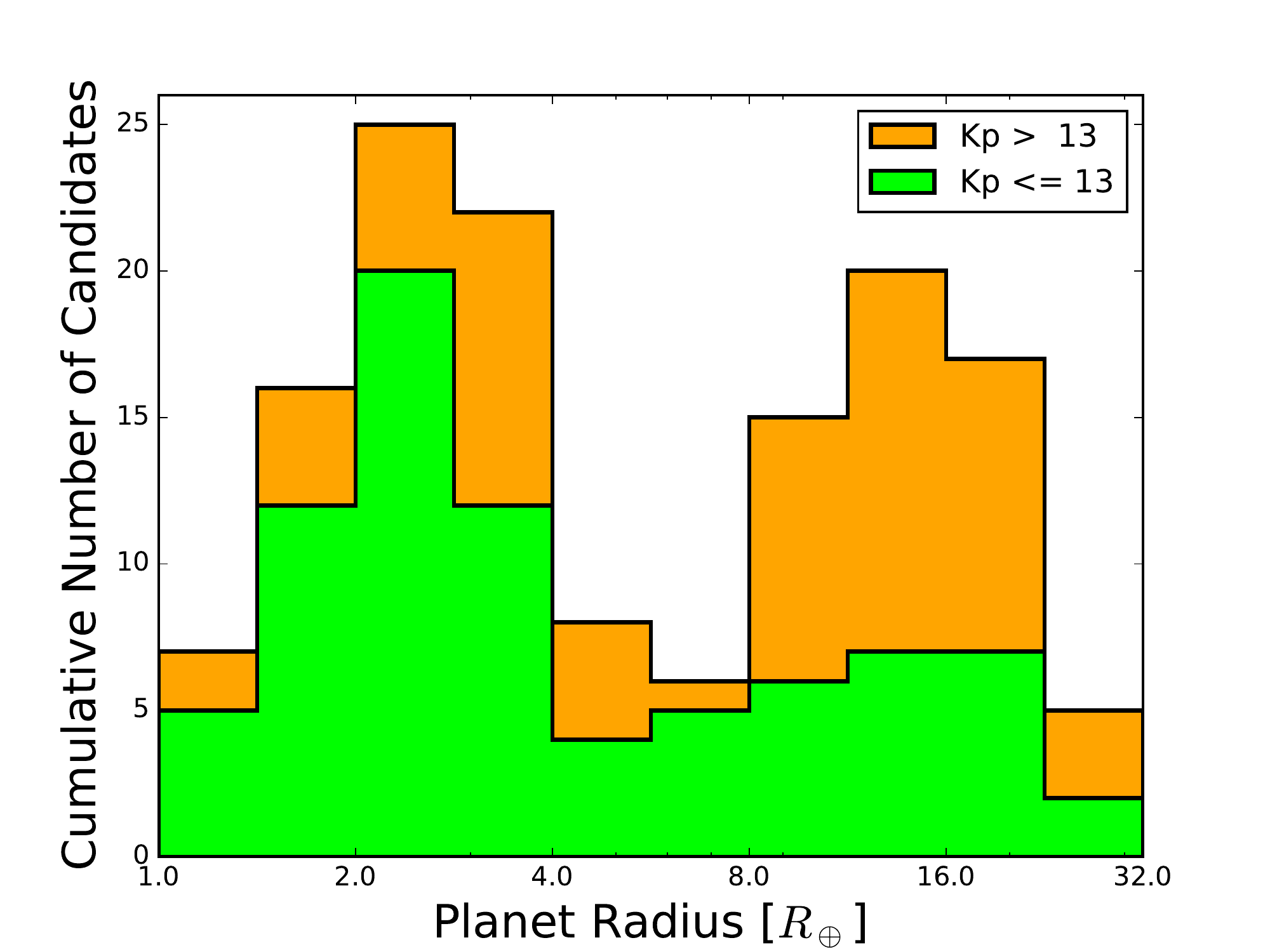}
\includegraphics[trim=0.5cm 0.5cm 1.5cm 0cm,width=0.32\textwidth]{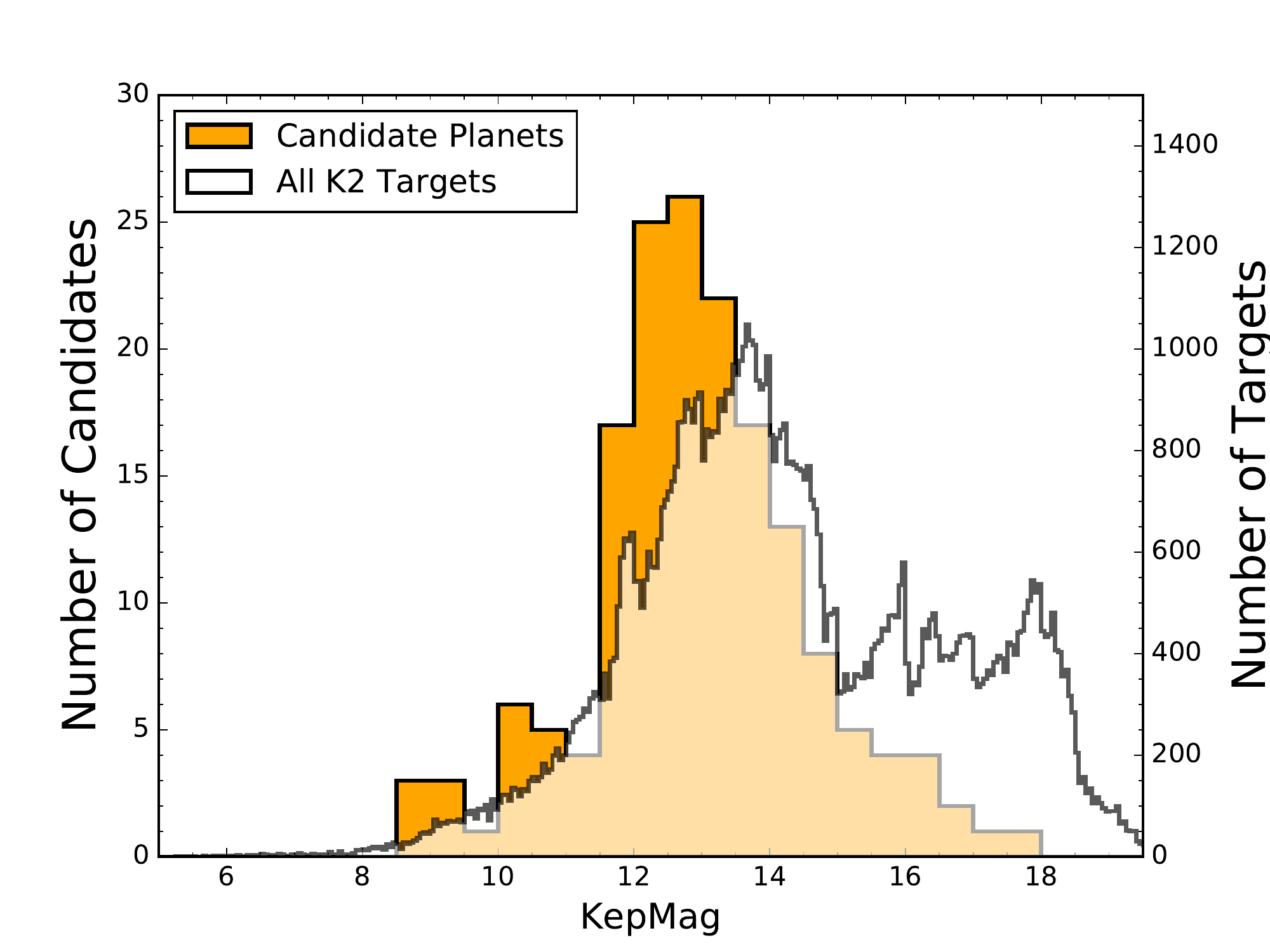}
\caption{{\em Left:} Orbital periods of transit-like signals identified in our analysis. The orange histogram (axis at left) indicates the distribution of planet candidates. The pale, narrow-binned histogram (axis at right) indicates the Threshold-Crossing Events (TCEs) identified by \terra in our initial transit search with MES\,$\ge10$.
{\em Middle:} Cumulative histograms of radii for our planet candidates. Most are moderately bright at $Kp \le 13$~mag, but at large radii over half orbit fainter stars; a large fraction of this second group are likely false positives. 
{\em Right:} The orange histogram (axis at left) shows the distribution of \kepmag for planet candidates. For comparison, the pale histogram (axis at right) shows all target stars from C5--C8.%
\label{fig:hist-per-prad-kepmag}}
\end{center}
\end{figure*}

Figure~\ref{fig:per-pradius} shows the distribution of the \stat{n-k2oi-not-eb} planet candidates in the $P-\Rp$ plane. Figure~\ref{fig:hist-per-prad-kepmag} shows 1D histograms of our candidates as a function of $P$, $\Rp$, and \kepmag. The median host star is nearly two magnitudes brighter in the \Kepler bandpass than the median KOI from the prime \Kepler mission (\stat{kepmag-k2oi-not-eb}~mag vs \stat{kepmag-koi}~mag, \citealt{Mullally15}). Our candidates have the following multiplicity distribution: \stat{n-k2oi-not-eb-stars-1} singles, \stat{n-k2oi-not-eb-stars-2} doubles, and \stat{n-k2oi-not-eb-stars-3} triple planet system.

We consulted the NASA Exoplanet Archive (NEA; \citealt{Akeson13})%
\footnote{https://exoplanetarchive.ipac.caltech.edu/}
to check whether previous analyses have reported significant numbers of candidates presented in this work. Of the catalogs incorporated into the NEA as of 2017-11-09, \citeauthor{Barros16} (2016; B16 hereafter) and  \citeauthor{Pope16} (2016; P16 hereafter) included 10 or more candidates from C5--C8.

B16 reported 172 planet candidates from C1--C6, of which 86 were in C5 and C6. Our catalog contains 107 candidates from C5 and C6. The two catalogs share 49 candidates. There are 58 candidates in our catalog that are not in B16, and there are 37 candidates in B16 that are not in our catalog. 

P16 reported 168 planet candidates in C5 and C6. Of these, our catalog includes 73 candidates. There are 34 candidates in our catalog that are not in P16, and there are 95 candidates in P16 that are not in our catalog.

\begin{figure*}[ht!]
\centering
\includegraphics[trim=1.75cm 1cm 1.75cm 0cm, width=0.32\textwidth]{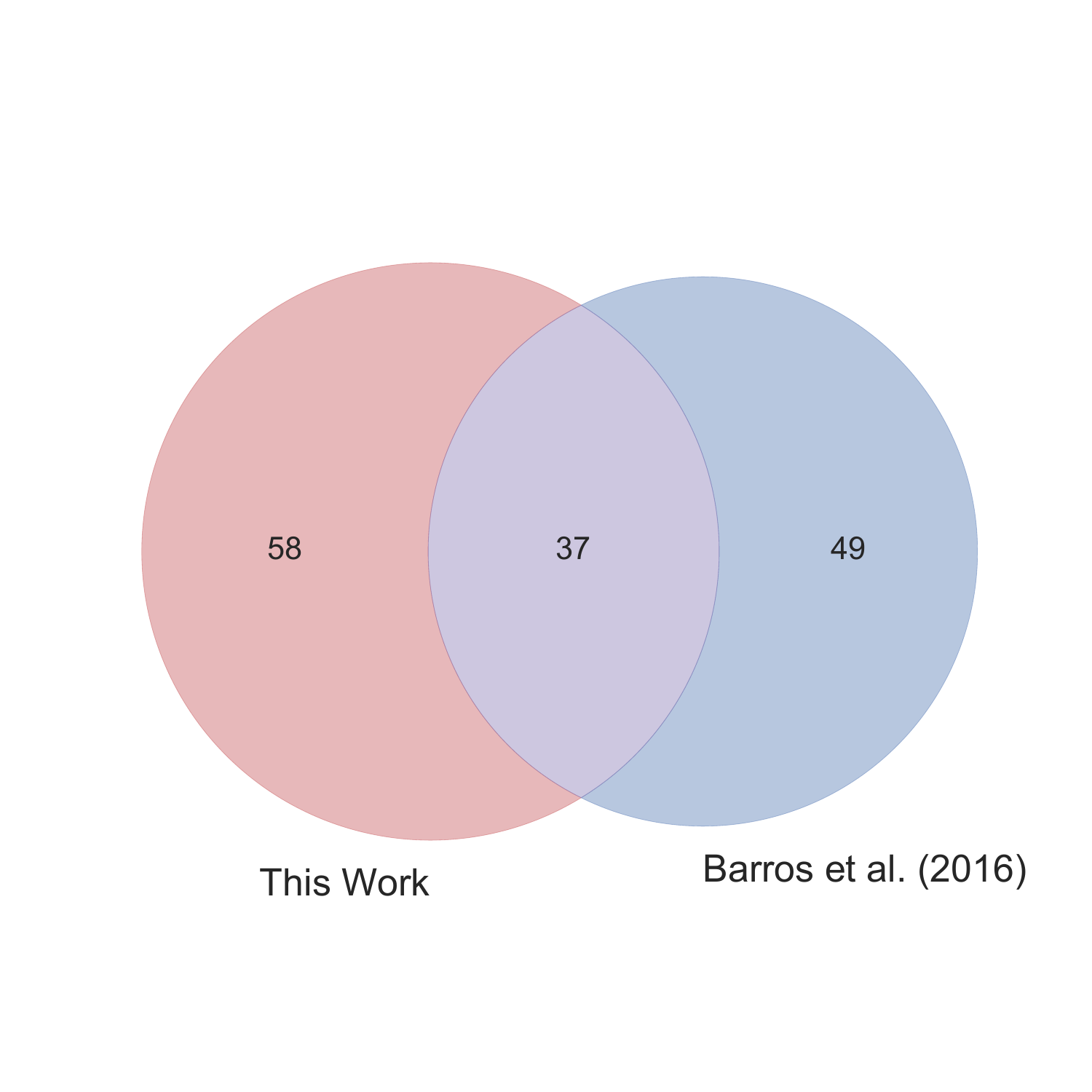}
\includegraphics[trim=1.75cm 1cm 1.75cm 0cm,width=0.32\textwidth]{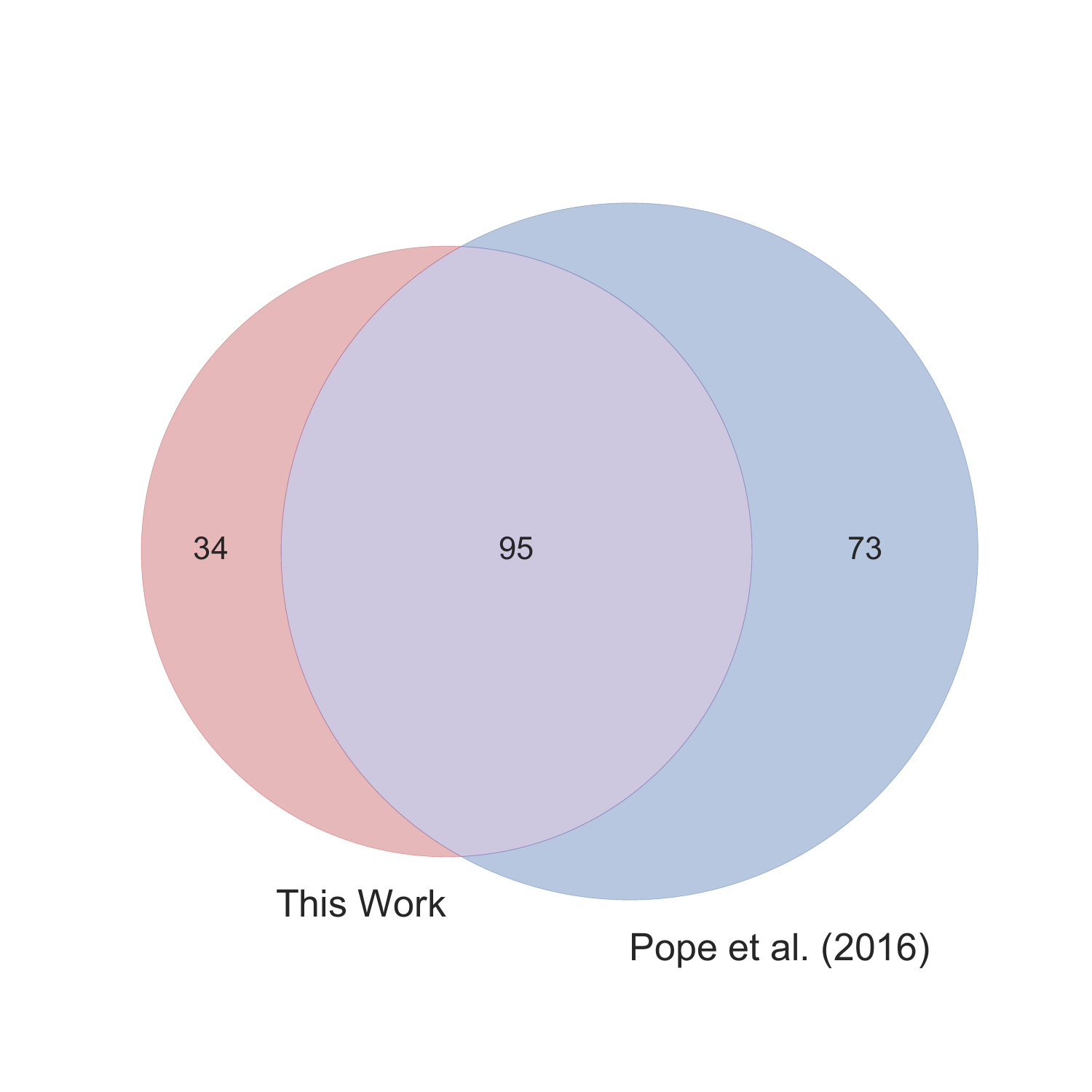}
\includegraphics[trim=1.75cm 1cm 1.75cm 0cm,width=0.32\textwidth]{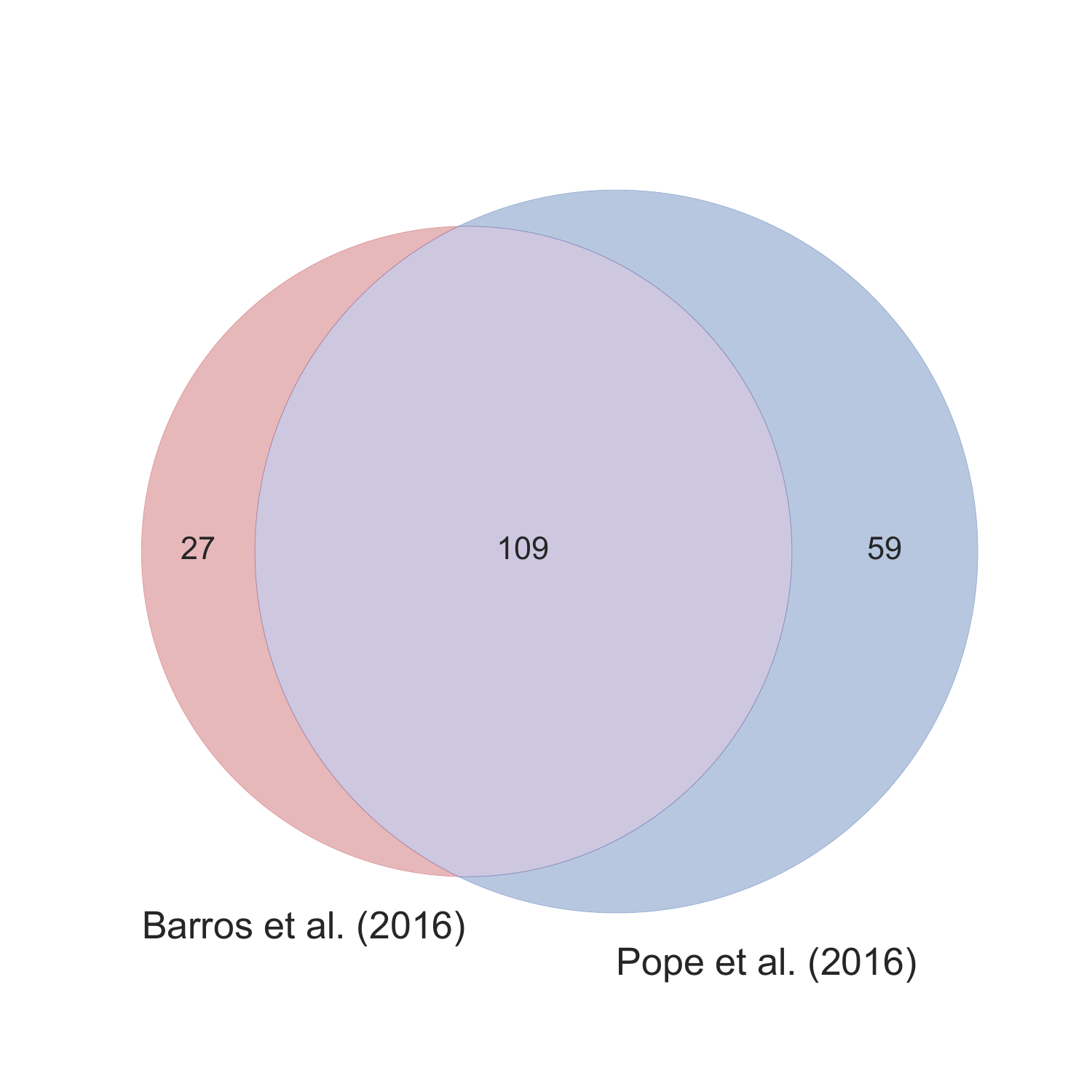}
\caption{Venn diagrams that compare the candidates from C5 and C6 reported in the following catalogs: This work, \cite{Barros16}, and \cite{Pope16}.\label{fig:venn}}
\end{figure*}

As a final point of comparison, B16 and P16 share 59 candidates. There are 27 candidates in B16 that are not in P16, and there are 109 candidates in P16 that were not in B16. Figure~\ref{fig:venn} is a Venn diagram that summarizes the degree of overlap between the various samples. 

A detailed analysis of why any particular candidate appeared in one catalog and not another is beyond the scope of this work. Broadly-speaking, the lack perfect overlap likely arises due to differences in photometric extraction algorithms, transit search algorithms, adopted signal-to-noise threshold for candidate status, and vetting diagnostics.

\section{Summary}
\label{sec:conclusions}

We report \stat{n-k2oi-not-eb} planet candidates orbiting \stat{n-k2oi-not-eb-stars} stars detected through a systematic search of \ktwo photometry from campaigns C5--C8. We also identified \stat{n-k2oi-eb} likely EBs based on their light curve morphology. We obtained Keck/HIRES optical spectra of \stat{n-hires-spec-pc}/\stat{n-k2oi-not-eb-stars} planet candidate host stars and \stat{n-hires-spec-eb}/\stat{n-k2oi-eb} EBs to improve our understanding of host star and planet properties and to search for binary companions. 

A small fraction of our planet candidates reside in multi-candidate systems (\stat{n-k2oi-not-eb-stars-2} doubles and \stat{n-k2oi-not-eb-stars-3} triple). These systems have a low false positive probability ($\lesssim 1\%$) due to their multiplicity \citep{Lissauer12}. The remaining \stat{n-k2oi-not-eb-stars-1} candidates are well-vetted and well-characterized planet candidates, but have yet been confirmed or statistically validated. Statistical validation requires a detailed analysis of light curve shape and constraints on the presence of blends from high-resolution imaging. \cite{Crossfield16} performed such an analysis to validate 104 K2 planet candidates identified during C0--C4. Our team's high contrast imaging followup will be presented in Gonzales et al. {\em in prep.} An analysis of the false positive probabilities of our candidates will be presented in Livingston et al. {\em in prep.}

Our typical candidate is two magnitudes brighter than the typical candidate from the \Kepler prime mission due to the larger region of sky observed by \ktwo. As a result, these candidates make up a valuable sample for further characterization efforts, such as Doppler measurements of planet masses.
 
\software{batman \citep{Kreidberg15}, SpecMatch-Syn \citep{Petigura15thesis}, SpecMatch-Emp \citep{Yee17}, k2phot (https://github.com/petigura/k2phot), isoclassify \citep{Huber17}, isochrones \citep{Morton15iso}.} 
 
\acknowledgments 
We thank the anonymous referee for a thoughtful reading of the manuscript and for useful suggestions. EAP acknowledges support from Hubble Fellowship grant HST-HF2-51365.001-A awarded by the Space Telescope Science Institute, which is operated by the Association of Universities for Research in Astronomy, Inc. for NASA under contract NAS 5-26555.  
Work by CDD was performed in part under contract with the Jet Propulsion Laboratory (JPL) funded by NASA through the Sagan Fellowship Program executed by the NASA Exoplanet Science Institute.
This research used the computing resources of NERSC, a DOE Office of Science User Facility supported by the Office of Science of the U.S. Department of Energy under Contract No. DE-AC02-05CH11231.
Finally, the authors wish to recognize and acknowledge the very significant cultural role and reverence that the summit of Maunakea has always had within the indigenous Hawaiian community.  We are most fortunate to have the opportunity to conduct observations from this mountain.

\bibliography{manuscript.bib}


\appendix
\begin{deluxetable*}{llRRlRhhRRrrrp{4.0cm}}
\tablecaption{Planet Candidates\label{tab:candidates}}
\tabletypesize{\scriptsize}
\tablecolumns{12}
\tablewidth{-5pt}
\tablehead{
	\multicolumn{2}{c}{Cand.} &
	\colhead{\kepmag} & 
	\colhead{\Rstar} & 
	\colhead{Prov.} & 
	\colhead{$P$} & 
	\nocolhead{$P$} & 
	\nocolhead{$T_{0}$} & 
	\colhead{\Rp/\Rstar} &
	\colhead{$T_{14}$} &
	\colhead{$b$} &
  	\colhead{\Rp} &
    \colhead{EB} &
	\colhead{Comments}\\
	\multicolumn{2}{c}{} &
	\colhead{mag} & 
	\colhead{\Rsun} & 
	\colhead{} & 
	\colhead{days} & 
	\nocolhead{days} & 
	\nocolhead{$\mathrm{BJD}_\mathrm{TDB}- 2454833$} & 
	\colhead{\%} &
	\colhead{hr} &
	\colhead{} &
  	\colhead{\Re} &
	\colhead{} &
	\colhead{}
	}
\startdata
C5 & 211401787.01 & 9.7 & 1.25 & S & 13.8 & ${ 13.772626 }^{ +0.001459 }_{ -0.001391 }$ & ${ 2318.069460 }^{ +0.003023 }_{ -0.003197 }$ & ${ 1.60 }^{ +0.13 }_{ -0.06 }$ & ${ 4.54 }^{ +0.13 }_{ -0.12 }$ & ${ 0.43 }^{ +0.32 }_{ -0.30 }$ & ${ 2.2 }^{ +0.4 }_{ -0.2 }$ & 0 & \\
C5 & 211945201.01 & 10.1 & 1.48 & S & 19.5 & ${ 19.492132 }^{ +0.000655 }_{ -0.000669 }$ & ${ 2325.824810 }^{ +0.000841 }_{ -0.000861 }$ & ${ 3.73 }^{ +0.23 }_{ -0.08 }$ & ${ 3.58 }^{ +0.12 }_{ -0.05 }$ & ${ 0.46 }^{ +0.28 }_{ -0.32 }$ & ${ 6.0 }^{ +0.9 }_{ -0.8 }$ & 0 & \\
C5 & 211990866.01 & 10.4 & 1.23 & S & 1.7 & ${ 1.673847 }^{ +0.000055 }_{ -0.000055 }$ & ${ 2307.720480 }^{ +0.001394 }_{ -0.001406 }$ & ${ 2.72 }^{ +0.26 }_{ -0.10 }$ & ${ 1.54 }^{ +0.05 }_{ -0.05 }$ & ${ 0.46 }^{ +0.32 }_{ -0.32 }$ & ${ 3.6 }^{ +0.4 }_{ -0.2 }$ & 0 & $\sim$1\% spot modulation\\
C5 & 212099230.01 & 10.5 & 0.96 & S & 7.1 & ${ 7.112273 }^{ +0.000284 }_{ -0.000278 }$ & ${ 2316.072880 }^{ +0.001427 }_{ -0.001488 }$ & ${ 3.02 }^{ +0.11 }_{ -0.10 }$ & ${ 3.13 }^{ +0.05 }_{ -0.06 }$ & ${ 0.97 }^{ +0.01 }_{ -0.01 }$ & ${ 3.2 }^{ +0.3 }_{ -0.2 }$ & 0 & \\
C5 & 211594205.01 & 10.7 & 0.79 & S & 17.0 & ${ 16.995848 }^{ +0.001351 }_{ -0.001361 }$ & ${ 2315.498660 }^{ +0.002431 }_{ -0.002323 }$ & ${ 1.82 }^{ +0.24 }_{ -0.10 }$ & ${ 2.49 }^{ +0.11 }_{ -0.09 }$ & ${ 0.48 }^{ +0.30 }_{ -0.33 }$ & ${ 1.6 }^{ +0.2 }_{ -0.1 }$ & 0 & \\
C5 & 212110888.01 & 11.4 & 1.38 & S & 3.0 & ${ 2.995633 }^{ +0.000008 }_{ -0.000008 }$ & ${ 2311.346840 }^{ +0.000098 }_{ -0.000097 }$ & ${ 8.70 }^{ +0.05 }_{ -0.05 }$ & ${ 2.36 }^{ +0.01 }_{ -0.02 }$ & ${ 0.77 }^{ +0.01 }_{ -0.01 }$ & ${ 13.1 }^{ +1.8 }_{ -1.6 }$ & 0 & \\
C5 & 211525389.01 & 11.7 & 0.98 & S & 8.3 & ${ 8.267084 }^{ +0.000258 }_{ -0.000254 }$ & ${ 2314.990350 }^{ +0.001175 }_{ -0.001156 }$ & ${ 3.28 }^{ +0.19 }_{ -0.07 }$ & ${ 3.40 }^{ +0.06 }_{ -0.05 }$ & ${ 0.38 }^{ +0.29 }_{ -0.27 }$ & ${ 3.5 }^{ +0.4 }_{ -0.2 }$ & 0 & \\
\enddata
\tablecomments{\ktwo objects of interest (K2OIs) identified through our systematic search of \ktwo photometry sorted by campaign and \Kepler magnitude (\kepmag). For each candidate we quote the highest quality stellar radius (\Rstar) computed from spectroscopy (S), where available, and from photometry (P), otherwise. Planet orbital period (\per{}), planet-to-star radius ratio (\rrat), time between first and last contact (\T{14}), and impact parameter ($b$) are provided. The full precision and uncertainties on $P$ and time of first transit $T_0$ are provided in the machine-readable version of this table. \Rp is the derived planetary radius given the light curve fit and stellar radius. We also include a summary of our eclipsing binary assessment: `EB' is a numerical code designating whether or the object of interest was a likely EB: 0---no obvious indication of EB, 1---secondary eclipse visible, 2---photometric variability that is phase-locked to the eclipse. Some additional comments on individual systems are provided.}
\end{deluxetable*}

\begin{deluxetable*}{lRRRRRRRRRRr}
\tablecaption{Stars with HIRES Spectra and Derived Parameters\label{tab:spec}}
\tabletypesize{\scriptsize}
\tablecolumns{12}
\tablewidth{0pt}
\tablehead{
	\colhead{Camp.} & 
	\colhead{EPIC} & 
	\colhead{\kepmag} & 
	\colhead{\teff} & 
	\colhead{\logg} & 
	\colhead{\fe} &
	\colhead{\vsini} &
    \colhead{\Mstar} & 
	\colhead{\Rstar} & 
	\colhead{SM\tablenotemark{a}} & 
	\colhead{SB2\tablenotemark{b}} &
    \colhead{disp\tablenotemark{c}} \\ 
	\colhead{} & 
	\colhead{} & 
	\colhead{mag} & 
	\colhead{K} & 
	\colhead{cgs} & 
	\colhead{dex} &
	\colhead{\kms} &
    \colhead{\Msun} & 
	\colhead{\Rsun} & 
	\colhead{} & 
	\colhead{} &
	\colhead{} 
	}
\startdata
C5 & 211401787 & 9.7 & 6214 & 4.26 & -0.04 & 8.4 & ${ 1.12 }^{ +0.06 }_{ -0.05 }$ & ${ 1.25 }^{ +0.17 }_{ -0.12 }$ & 1 & 1 & PC\\
C5 & 211945201 & 10.1 & 6018 & 4.13 & 0.12 & 3.3 & ${ 1.17 }^{ +0.10 }_{ -0.08 }$ & ${ 1.48 }^{ +0.21 }_{ -0.19 }$ & 1 & 1 & PC\\
C5 & 211990866 & 10.4 & 6180 & 4.51 & 0.32 & 13.6 & ${ 1.25 }^{ +0.05 }_{ -0.04 }$ & ${ 1.23 }^{ +0.09 }_{ -0.06 }$ & 1 & 1 & PC\\
C5 & 212099230 & 10.5 & 5487 & 4.41 & 0.11 & 1.6 & ${ 0.93 }^{ +0.04 }_{ -0.04 }$ & ${ 0.96 }^{ +0.09 }_{ -0.06 }$ & 1 & 1 & PC\\
C5 & 211594205 & 10.7 & 5240 & 4.69 & -0.05 & 1.5 & ${ 0.83 }^{ +0.03 }_{ -0.03 }$ & ${ 0.79 }^{ +0.04 }_{ -0.03 }$ & 1 & 1 & PC\\
C5 & 212110888 & 11.4 & 6008 & 4.16 & 0.01 & 5.9 & ${ 1.09 }^{ +0.09 }_{ -0.06 }$ & ${ 1.38 }^{ +0.19 }_{ -0.17 }$ & 1 & 4 & PC\\
C5 & 211525389 & 11.7 & 5479 & 4.48 & 0.30 & 2.4 & ${ 1.00 }^{ +0.04 }_{ -0.04 }$ & ${ 0.98 }^{ +0.08 }_{ -0.05 }$ & 1 & 1 & PC\\
C5 & 211359660 & 11.7 & 5177 & 4.62 & 0.12 & 2.0 & ${ 0.87 }^{ +0.03 }_{ -0.03 }$ & ${ 0.82 }^{ +0.04 }_{ -0.03 }$ & 1 & 1 & PC\\
\enddata
\tablecomments{The $\stat{n-hires-spec}$ stars observed by Keck/HIRES, their stellar parameters, and SB2 status (if available). We derived stellar properties using two complimentary techniques: \SpecMatchSyn \citep{Petigura15thesis} and \SpecMatchEmp \citep{Yee17}. For stars cooler than 4700~K, we used \SpecMatchSyn to compute \teff, \logg, and \fe, which were converted into \Rstar and \Mstar using the \isochrones package \citep{Morton15iso}. For stars cooler than 4700~K, we used \SpecMatchEmp to compute \teff, \Rstar, \Mstar, and \fe. We also searched for SB2s using the methodology of \cite{Kolbl15}.}
\tablenotetext{a}{Source of spectroscopic properties: 0---Spectroscopic properties are not reliable ($\stat{n-sm-none}$ stars). 1---Parameters from \SpecMatchSyn ($\stat{n-sm-smsyn}$ stars). 2---Parameters from \SpecMatchEmp ($\stat{n-sm-smemp}$ stars). The spectroscopic parameters are not reliable if \teff > 6500~K or \vsini > 20~\kms. \SpecMatchSyn achieves precisions of 60~K, 0.10~dex, 0.04~dex, and 1~\kms in \teff, \logg, \fe, and \vsini. \SpecMatchEmp achieves precisions of 70~K in \teff, 10\% in \Rstar, and 0.12~dex in \fe.}
\tablenotetext{b}{\ReaMatch classification codes: 1---No detection, 2---Star is unfit for \ReaMatch: Teff below 3500 K or above 6500K, 3---Star is unfit for \ReaMatch: \vsini above 10~\kms; 4---Ambiguous detection; 5---Obvious detection.}
\tablenotetext{c}{Flag indicating whether the target is a planet candidate (PC), eclipsing binary (EB), or is not included in our candidate list (Other).}
\end{deluxetable*}
\end{document}